\documentclass{article}
\usepackage[preprint]{spconfa4}
\usepackage{amsmath,graphicx}

\usepackage{amssymb}
\usepackage{enumerate}
\usepackage{comment}
\usepackage{color}
\usepackage{bm}
\usepackage{array}
\usepackage{multirow}
\usepackage{colortbl}
\usepackage{url}

\newlength\savedwidth
\newcommand{\wcline}[1]{\noalign{\global\savedwidth\arrayrulewidth\global\arrayrulewidth 1.0pt} \cline{#1}
\noalign{\global\arrayrulewidth\savedwidth}}

\copyrightnotice{978-1-6654-6867-1/22/\$31.00~\copyright2022 IEEE}
                   
\title{Joint Analysis of Acoustic Scenes and Sound Events\\with Weakly labeled Data}
\name{Shunsuke Tsubaki$^{\hspace{1pt} \ast \dagger}$, Keisuke Imoto$^{\hspace{1pt} \ast \dagger}$, and Nobutaka Ono$^{\hspace{1pt} \ddagger}$\thanks{$^\ast$These authors contributed equally to this work.}}
\address{$^\dagger$Doshisha University, Japan, $^\ddagger$Tokyo Metropolitan University, Japan}
\begin{document}
%
\maketitle
%
\begin{abstract}
Considering that acoustic scenes and sound events are closely related to each other, in some previous papers, a joint analysis of acoustic scenes and sound events utilizing multitask learning (MTL)-based neural networks was proposed. In conventional methods, a strongly supervised scheme is applied to sound event detection in MTL models, which requires strong labels of sound events in model training; however, annotating strong event labels is quite time-consuming. In this paper, we thus propose a method for the joint analysis of acoustic scenes and sound events based on the MTL framework with weak labels of sound events. In particular, in the proposed method, we introduce the multiple-instance learning scheme for weakly supervised training of sound event detection and evaluate four pooling functions, namely, max pooling, average pooling, exponential softmax pooling, and attention pooling. Experimental results obtained using parts of the TUT Acoustic Scenes 2016/2017 and TUT Sound Events 2016/2017 datasets show that the proposed MTL-based method with weak labels outperforms the conventional single-task-based scene classification and event detection models with weak labels in terms of both the scene classification and event detection performances.
\end{abstract}
%
%
\section{Introduction}
\label{sec:intro}
%
Environmental sound analysis (ESA), in which sounds that are not limited to voice or music are analyzed, has attracted considerable attention in recent years.
ESA is expected to have various applications such as media retrieval \cite{Fonseca_DCASE2018_01}, automatic surveillance \cite{Chan_EUSIPCO2010_01}, machine condition monitoring \cite{Koizumi_DCASE2020_01}, and biomonitoring systems \cite{Morfi_JASA2021_01}.
In ESA, acoustic scene classification (ASC) and sound event detection (SED) are the two major tasks addressed.
In the former ESA, a predefined acoustic scene label is predicted from an audio recording, where the acoustic scene indicates the place, situation, or surrounding in which the audio is recorded, such as ``office,'' ``meeting,'' or ``indoor.''
On the other hand, SED involves the detection of sound event labels and their time boundaries in the audio recording, where a sound event indicates a sound class, such as ``footstep,'' ``car,'' or ``cutlery.''

In recent years, many researchers have applied neural-network-based methods to ASC and SED.
For instance, Valenti et al. \cite{Valenti_IJCNN2017_01} and Raveh and Amar \cite{Raveh_DCASE2018_01} proposed convolutional neural-network (CNN)- and ResNet-based ASC methods, respectively.
\c{C}ak\i r et al. proposed an event detection method using a convolutional recurrent neural network (CRNN), which can capture temporal information of sound events \cite{Cakir_TASLP2017_01}.
More recently, Kong et al. \cite{Kong_TASLP2020_01} have proposed SED methods utilizing the Transformer encoder.
In most conventional methods, acoustic scenes and sound events are analyzed separately.
On the other hand, acoustic scenes and sound events are closely related.
For instance, in the acoustic scene ``home,'' the sound events ``cutlery'' and ``glass jingling'' tend to occur, whereas the sound events ``wind blowing'' and ``large vehicle'' occur infrequently.
On the basis of this idea, Mesaros et al. \cite{Mesaros_EUSIPCO2011_01} and Heittola et al. \cite{Heittola_JASM2013_01} proposed methods utilizing information on acoustic scenes for SED.
Imoto and co-workers proposed ASC methods considering the relationship between sound events and acoustic scenes using Bayesian generative models \cite{Imoto_IEICE2016_01,Imoto_TASLP2019_01}.
Bear et al. \cite{Bear_INTERSPEECH2019_01}, Tonami et al. \cite{Tonami_WASPAA2019_01}, and Jung et al. \cite{Jung_ICASSP2021_01} proposed methods for the joint analysis of acoustic scenes and sound events utilizing the MTL-based neural network models of ASC and SED.
These methods require a large amount of strongly annotated data, which has sound event classes and their timestamp.
However, annotating strong labels to a large amount of sound data is quite time-consuming.
To address this problem, many researchers have proposed weakly supervised SED, which only requires weakly labeled data in the model training stage \cite{Kumar_ACMMM2016_01,Turpault_DCASE2019_01,Zheng_DCASE2021_01}.
The weakly labeled data has only tag information for each sound clip and is relatively easy to be collected.
In this work, we thus introduce the weakly supervised training scheme in the joint analysis of acoustic scenes and sound events using the MTL framework.

%
\begin{figure*}[t!]
\centering
\hspace*{-5pt}
\begin{tabular}{c}
\begin{minipage}{0.40\linewidth}
\centering
\vspace{-5pt}
\includegraphics[width=0.90\columnwidth]{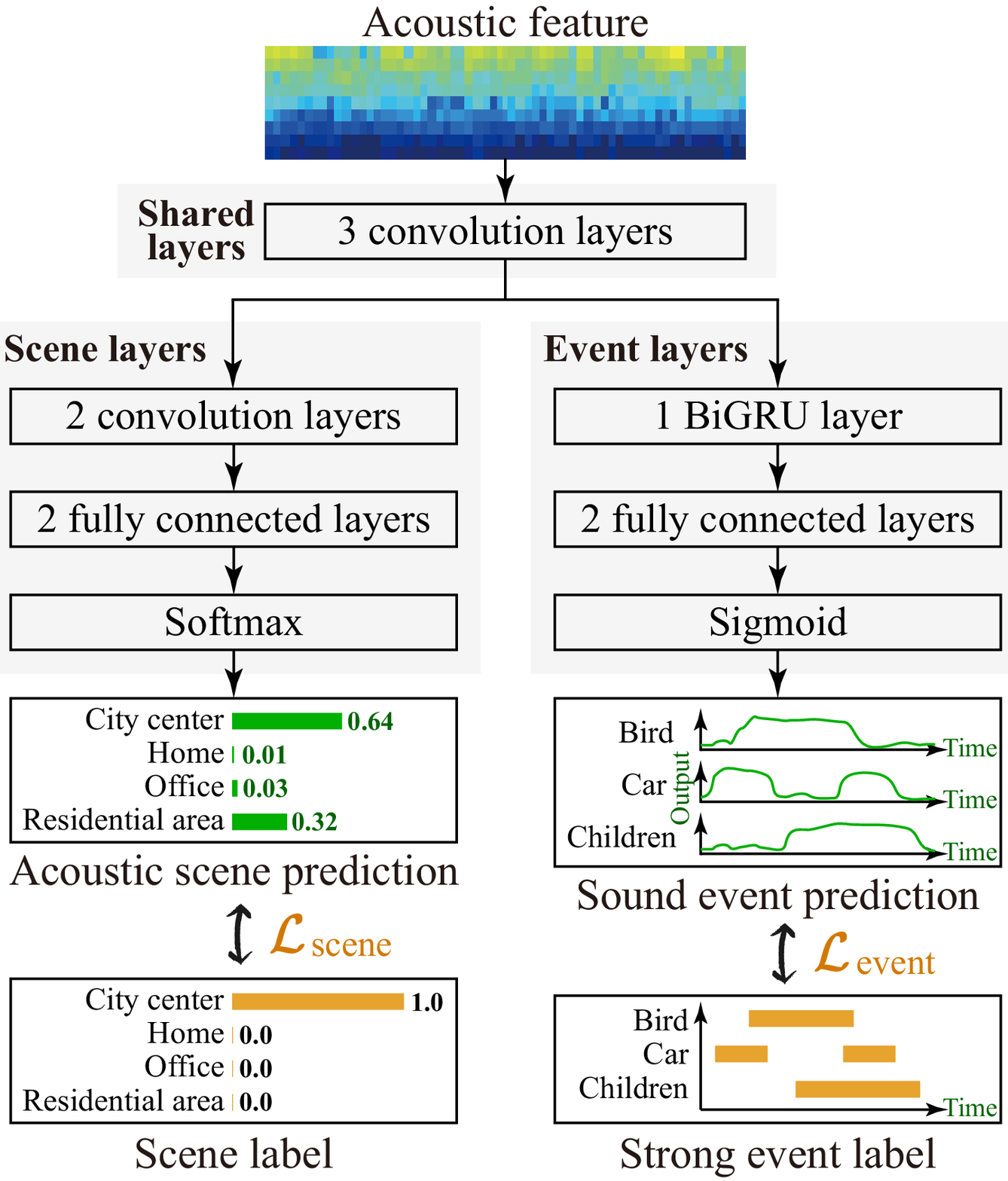}
\vspace{0pt}
\caption{Network structure of conventional MTL-based method \cite{Tonami_WASPAA2019_01}}
\label{fig:conventionalMTL}
\end{minipage}
\hspace{11pt}
\begin{minipage}{0.57\linewidth}
\vspace{-10pt}
\centering
\includegraphics[width=0.94\columnwidth]{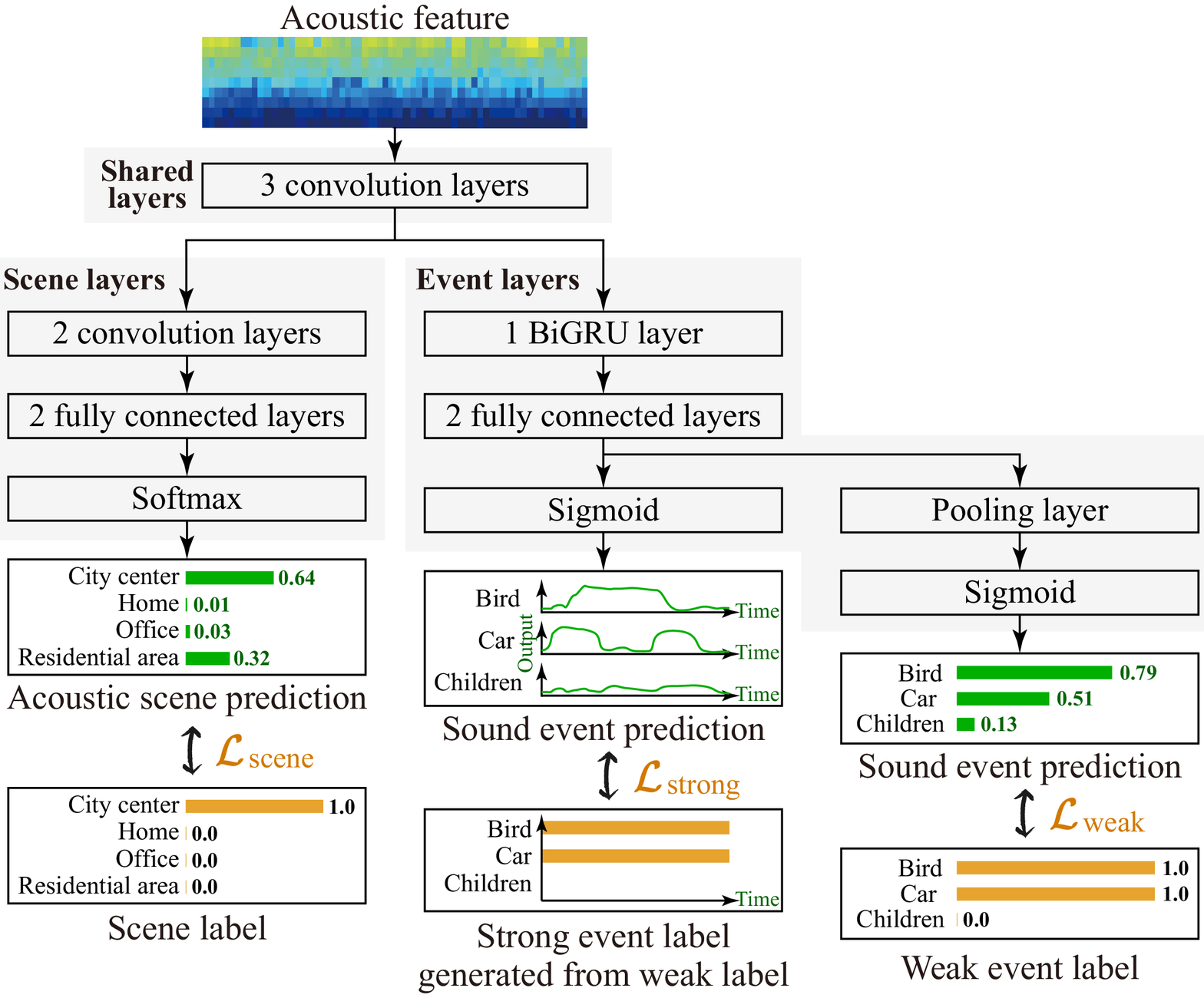}
\vspace{-6pt}
\caption{Network structure of proposed MTL of ASC and SED using weak labels of sound events}
\label{fig:proposedMTL}
\end{minipage}
\end{tabular}
\vspace{5pt}
\end{figure*}

The rest of this paper is organized as follows.
In section 2, we introduce the conventional scene classification and event detection methods, and the joint analysis of scenes and events based on the MTL framework.
In section 3, we propose the method for the joint analysis of acoustic scenes and sound events with weak labels of sound events.
In section 4, we describe our experiment conducted to evaluate the ASC and SED performances of the proposed method.
Finally, we conclude this work in section 5.
%
%
\section{Conventional Methods}
\label{sec:conventional}
\vspace{-3pt}
\subsection{Conventional Scene Classification and Event Detection Methods}
\label{ssec:ConvASCandSED}
Many conventional ASC and SED methods apply neural-network-based models such as CNN \cite{Valenti_IJCNN2017_01,Hershey_ICASSP2017_01}, CRNN \cite{Cakir_TASLP2017_01}, and Transformer encoder \cite{Kong_TASLP2020_01}.
We here overview the conventional ASC and SED methods using neural networks.
As the acoustic feature for ASC and SED, many systems use the time--frequency representation of the acoustic signal $X \in \mathcal{R}^{D \times T}$, such as the time series of mel frequency cepstrum coefficients (MFCCs) or the log mel-band spectrogram.
Here, $D$ and $T$ are the numbers of frequency bins and time frames, respectively.
The time--frequency representation of an acoustic signal is fed to the ASC or SED network.
In ASC, the model parameters are trained using the network output and the cross-entropy (CE) loss function ${\mathcal L}_{{\rm scene}}$,

\vspace{-9pt}
\begin{align}
{\mathcal L}_{{\rm scene}} &= - \! \sum^{N}_{n=1} {\Big \{} z_{n} \log ( y_{n} ) {\Big \}},
\label{eq:scene_loss}
\end{align}
\vspace{-4pt}

\noindent where $N$, $y_{n}$, and $z_{n}$ are the number of acoustic scene classes, the prediction result, and the target scene label, respectively.
The target scene label that the acoustic signal is most associated with is 1 and 0 otherwise.

The model parameters of SED are trained using the network output and the following binary cross-entropy (BCE) loss function ${\mathcal L}_{{\rm event}}$,

\vspace{-6pt}
\begin{align}
\ \nonumber\\[-38pt]
\hspace{0pt} {\mathcal L}_{{\rm event}} &= - \! \sum^{T}_{t=1} \hspace{-1pt} {\Big \{} {\bf z}_{t} \log ( {\bf y}_{t} ) \! + \! (1 \! - \! {\bf z}_{t}) \log (1 \! - \! {\bf y}_{t} ) \hspace{-1pt} {\Big \}} \nonumber\\[1pt]
&\hspace{0pt} = - \!\!\! \sum^{T\!, \hspace{1pt} M}_{t,m=1} \hspace{-4pt} {\Big \{} \hspace{-0.5pt} z_{t,m} \log ( y_{t,m} ) \! + \! (1 \! - \! z_{t,m}) \log (1 \! - \! y_{t,m}) \hspace{-1.2pt} {\Big \}} \hspace{-0.3pt} ,\hspace{-4pt}
\label{eq:event_loss}
\end{align}
\vspace{-4pt}

\noindent where $T$, $M$, $y_{t,m}$, and $z_{t,m}$ are the number of time frames, the number of sound event classes, the prediction of sound event $m$ in time frame $t$, and the target event label, respectively.
%
%
\subsection{Joint Analysis of Acoustic Scenes and Sound Events Based on Multitask Learning}
\label{ssec:ConvMTL}
In most conventional methods, acoustic scenes and sound events are analyzed separately.
However, many scenes and events are closely related, and information on acoustic scenes and sound events is beneficial for SED and ASC, respectively.
Thus, the method for the joint analysis of acoustic scenes and sound events based on the multitask learning (MTL) of ASC and SED has been proposed \cite{Bear_INTERSPEECH2019_01,Tonami_WASPAA2019_01,Jung_ICASSP2021_01,Imoto_ICASSP2020_01}.

As shown in Fig.~\ref{fig:conventionalMTL}, conventional methods share the part of the network that holds information on acoustic scenes and sound events in the shared layers.
The CNN and recurrent neural network (RNN) layers are applied to scene classification and event detection networks, respectively.
To train the MTL model of ASC and SED, the conventional methods apply the following loss function:

\vspace{-8pt}
\begin{align}
{\mathcal L} &= \alpha {\mathcal L}_{{\rm scene}} + \beta {\mathcal L}_{{\rm event}},
\label{eq:mtl_loss}
\end{align}
\vspace{-13pt}

\noindent where $\alpha$ and $\beta$ are the constant weights of scene classification and event detection losses, respectively.
In this work, $\beta = 1.0$ is set without loss of generality.
\begin{table}[t]
\vspace{0pt}
\small
\caption{Pooling functions used for proposed method}
\vspace{2pt}
\label{tab:pooling}
\begin{center}
\begin{tabular}{ll}
\wcline{1-2}
&\\[-9pt]
\multicolumn{1}{c}{\textbf{Pooling function}} & \multicolumn{1}{c}{\ \ \textbf{Forward}} \\
\wcline{1-2}
&\\[-8.0pt]
\textbf{Max pooling (MP)} &\ \  $y = \max x_{t}$ \\
&\\[-8.0pt]
\textbf{Average pooling (AP)} &\ \  $\displaystyle y = \frac{1}{T} \sum_{t=1}^{T} x_{t}$ \\
&\\[-6.0pt]
\textbf{Exponential softmax pooling (ES)} \ \ &\ \  $\displaystyle y = \frac{\sum_{t} x_{t} \exp (x_{t})}{\sum_{t} \exp (x_{t})} $ \\[-2pt]
&\\[-2.0pt]
\textbf{Attention pooling (AT)} &\ \  $\displaystyle y = \frac{\sum_{t} w_{t} x_{t}}{\sum_{t} w_{t}}$ \\
&\\[-8.0pt]
\wcline{1-2}
\end{tabular}
\vspace{-4pt}
\end{center}
\end{table}
%
%
%
\vspace{-2pt}
\section{Proposed Method}
\label{sec:proposed}
\vspace{-3pt}
The conventional MTL-based joint analysis of acoustic scenes and sound events requires strong labels of sound events.
However, annotating strong labels of sound events is very time-consuming.
To address this limitation of conventional methods, we propose a weakly supervised method for the MTL of ASC and SED using weak labels of sound events.

In the proposed method, we apply the multiple-instance learning (MIL) framework \cite{Dietterich_AI1997_01} to the weakly supervised training of SED.
Figure~\ref{fig:proposedMTL} shows the network structure of the proposed MTL of ASC and SED using weak labels of sound events.
As shown in Fig.~\ref{fig:proposedMTL}, the proposed method has the shared, scene-specific, and event-specific layers as in the conventional MTL-based method, whereas the event-specific layers have two branches.
One branch has the sigmoid function after the recurrent and fully connected layers and is used to predict sound events and their timestamps as in the conventional methods.
The other branch has the MIL-based pooling layer and sigmoid function, which enable the training of a SED model with weak labels of sound events.
In particular, we apply four pooling functions for MIL, namely, max pooling (MP), average pooling (AP), exponential softmax pooling (ES), and attention pooling (AT) \cite{Wang_ICASSP2019_01}.
Table~\ref{tab:pooling} shows the definitions of pooling functions.

The model parameters of the proposed method are trained using Eq.~(\ref{eq:mtl_loss}) as in the conventional method, whereas the loss function $\mathcal{L}_{\rm event}$ is represented as

\vspace{-7pt}
\begin{align}
&\hspace{-2pt} \mathcal{L}_{\rm event} = \gamma \mathcal{L}_{\rm strong} + \zeta \mathcal{L}_{\rm weak}\nonumber\\[2pt]
&\hspace{2pt} = - \gamma \hspace{-1pt} \sum_{m,t=1}^{M\hspace{-0.8pt}, \hspace{0.8pt}T} \hspace{-2pt} {\Big \{} z_{m} \log s(y_{m,t}) \hspace{-1pt} + \hspace{-1pt} (1 \hspace{-1pt} - \hspace{-1pt} z_{m}) \log {\big (} 1 \hspace{-1pt} - \hspace{-1pt} s(y_{m,t}) {\big )} \hspace{-1pt} {\Big \}}\nonumber\\[0pt]
&\hspace{12pt} - \zeta \hspace{-1pt} \sum_{m=1}^{M} \hspace{-2pt} {\Big \{} z_{m} \log s(y_{m}) \hspace{-1pt} + \hspace{-1pt} (1-z_{m}) \log {\big (} 1 \hspace{-1pt} - \hspace{-1pt} s(y_{m}) {\big )} \hspace{-1pt} {\Big \}},\nonumber\\[-6pt]
\end{align}
\newpage

\noindent where $y_{m}$ and $z_{m}$ are the bag level prediction of sound event $m$ and the weak label of the target event, respectively.
$\gamma$ and $\zeta$ are the constant weights of event detection losses.
\begin{table}[t]
\small
\caption{Detailed structure of MTL network of ASC and SED}
\vspace{-17pt}
\label{tbl:networks}
\begin{center}
\begin{tabular}{ccc}
\wcline{1-3}
\!\!&\!\!\\[-9.5pt]
\multicolumn{3}{c}{\textbf{Shared layers}}\\
\wcline{1-3}
\!\!&\!\!\\[-9pt]
\multicolumn{3}{c}{Log-mel energy (500 frames $\times$ 64 mel bin)}\\[0pt]
\cline{1-3}
\!&\\[-9pt]
\multicolumn{3}{c}{3$\times$3 kernel size/128 ch.}\\[-1pt]
\multicolumn{3}{c}{Batch norm., Leaky ReLU}\\[-1pt]
\multicolumn{3}{c}{1$\times$8 Max pooling}\\[0pt]
\cline{1-3}
\!&\\[-9pt]
\multicolumn{3}{c}{$\begin{pmatrix} \textrm{3$\times$3 kernel size/128 ch.}\\[-1pt]
\textrm{Batch norm., Leaky ReLU}\\[-1pt]
\textrm{1$\times$2 Max pooling}
\end{pmatrix}$ $\times$ 2
}\\[0pt]
\!&\\[-9pt]
\wcline{1-3}
\multicolumn{1}{c}{}\\[-9.5pt]
\multicolumn{1}{c}{\textbf{Scene layers}}&\!\!&\textbf{Event layers}\\
\wcline{1-3}
\multicolumn{1}{c}{}\\[-9pt]
\multicolumn{1}{c}{3$\times$3 kernel size/256 ch.}&\!\!&\\[-1pt]
\multicolumn{1}{c}{Batch norm., Leaky ReLU}&\!\!&BiGRU w/ 32 units\\[-1pt]
\multicolumn{1}{c}{25$\times$1 Max pooling}&\!\!&\\
\cline{1-1} \cline{3-3}
\multicolumn{1}{c}{}\\[-9pt]
\multicolumn{1}{c}{3$\times$3 kernel size/256 ch.}&\!\!&\multirow{2}{*}{}\\[-1pt]
\multicolumn{1}{c}{Batch norm., Leaky ReLU}&\!\!&FC w/ 32 units, Leaky ReLU\\[-1pt]
\multicolumn{1}{c}{Global max pooling}&\!\!&\\
\cline{1-1} \cline{3-3}
\multicolumn{1}{c}{}\\[-9pt]
\multicolumn{1}{c}{FC w/ 32 units, Leaky ReLU}&\!\!&FC w/ 25 units, Sigmoid\\[-1pt]
\cline{1-1}\wcline{3-3}
\multicolumn{1}{c}{}\\[-9pt]
\multicolumn{1}{c}{FC w/ 4 units, Softmax}&\!\!&\\
\wcline{1-1}
\end{tabular}
\end{center}
\end{table}
\begin{table}[t]
\vspace{-13pt}
\small
\caption{Experimental conditions}
\vspace{2pt}
\label{tbl:parameter}
\begin{center}
\begin{tabular}{ll}
\wcline{1-2}
&\\[-8.5pt]
Acoustic feature & Log-mel energy (64 dim.)\\
Frame length/shift & 40 ms/20 ms\\
Length of sound clip & 10 s\\
Optimizer & RAdam \cite{Liu_ICLR2020_01}\\[0pt]
$\alpha$, $\beta$, $\gamma$, $\zeta$ & 0.001, 1.0, 0.1--1.0, 0.01--0.1\\[0pt]
\wcline{1-2}
\end{tabular}
\vspace{-5pt}
\end{center}
\end{table}
%
%
%
\section{Evaluation Experiments}
\label{sec:experiment}
\vspace{-3pt}
\subsection{Experimental Conditions}
\label{ssec:condition}
To evaluate the performances of ASC and SED using the proposed weakly supervised MTL method, we conducted the evaluation experiments using datasets composed of parts of the TUT Acoustic Scenes 2016/2017 and TUT Sound Events 2016/2017 \cite{Mesaros_EUSIPCO2016_01,Mesaros_DCASE2017_01}.
From these datasets, we selected acoustic signals including four acoustic scenes, ``city center,'' ``home,'' ``office,'' and ``residential area,'' which comprise a total of 266 min of sounds (192 min for the development set and 74 min for the evaluation set).
These sound clips include 25 types of sound event.
The details of the dataset used for the evaluation experiments are found in \cite{Imoto_dataset2019_01}.

For an acoustic feature, we used the 64-dimensional log mel-band energy, which was extracted with a 40 ms frame length and a 20 ms shift size.
As the ASC and SED models, we applied the MTL-based network as shown in Table~\ref{tbl:networks}, which has been proposed in \cite{Tonami_WASPAA2019_01}.
For each method, we conducted the evaluation experiments 10 times with random initial values of model parameters.
Other experimental conditions are listed in Table~\ref{tbl:parameter}.
%
%
\vspace{-5pt}
\subsection{Experimental Results}
\label{ssec:result}
\vspace{-3pt}
\subsubsection{Overall Performances of ASC and SED}
\label{sssec:result1}
\vspace{-2pt}
Table~\ref{tbl:performance01} shows the average performances of ASC and SED.
CNN (ASC) and CNN-BiGRU (SED) indicate the scene classification and event detection methods using single-task networks with the shared + scene and shared + event layers, respectively.
In this experiment, we evaluated the SED performance with the frame-based metric because the SED system outputs the frame-wise predictions, and we can understand the basic trends of the system output with this metric.
The results show that the proposed MTL-based method with weak event labels outperforms the single-task models in terms of both the scene classification and event detection performances.
With the proposed MTL and the max pooling layer, the micro- and macro-Fscores for SED are improved by 2.35 and 2.30 percentage points compared with those of CNN-BiGRU (SED), respectively.
Thus, even when we use the weak labels of sound events, information on acoustic scenes and sound events improves SED and ASC, respectively.

Compared with the conventional MTL with strong labels, the proposed method using weak labels of sound events achieves competitive performance in ASC.
This result implies that in ASC, the clip-level event labels are informative, whereas the timestamps are not necessarily required.
In SED, the conventional method using the strong labels shows a higher performance in terms of the micro-Fscore than the proposed method, which is not bad considering that the proposed method uses only weak labels of sound events.
In terms of the macro-Fscore, the proposed method outperforms the conventional method.
This is because the conventional method using strong labels is very conservative and hardly detects sound events with short durations.
On the other hand, since the weakly supervised method does not use the timestamp in the model training and depends less on the sound duration, the proposed method tends to detect sound events with short durations well.
For a detailed discussion on the reason for theses observations, refer to \cite{Imoto_ICASSP2021_01}.
%
%
\vspace{-7pt}
\subsubsection{Comparison of Pooling Functions for MIL}
\label{sssec:result2}
\vspace{-4pt}
We then compared the SED performances in terms of the four pooling functions.
Table~\ref{tbl:performance01} shows that the attention pooling outperforms the other pooling functions in terms of the overall performance.
To investigate how each pooling function affects the SED performance of the proposed method, we list the average Fscores for selected sound events in Table~\ref{tab:performance02}.
The results show that the max pooling is effective for detecting instantaneous sound events such as ``cutlery'' and ``keyboard typing,'' as well as easily distinguishable sound events from short frames such as ``bird singing.''
The results also indicate that the average pooling performs well for long-lasting sounds such as ``large vehicle.''
On the other hand, attention pooling can balance a reasonable performance between instantaneous and long-lasting sounds because it flexibly handles the sound events with various time lengths using the trainable focusing weights.
\begin{table}[t]
\vspace{-9pt}
\small
\caption{Overall performances of ASC and SED}
\vspace{0pt}
\label{tbl:performance01}
\centering
\begin{tabular}{lccccccc}
\wcline{1-6}
&\\[-9.5pt]
&\multicolumn{2}{c}{\textbf{Scene}}&\!\!\!&\multicolumn{2}{c}{\textbf{Event}} \!\!\\[-0.5pt]
\cline{2-3}\cline{5-6}
&\\[-9.5pt]
\multicolumn{1}{c}{\textbf{Method}}&\!\!\!\! Micro- \!\!\!\!&\!\!\!\! Macro- \!\!\!\!&\!\!\!& Micro- \!\!\!\!& Macro- \!\!\\[-1.5pt]
\multicolumn{1}{c}{} \!\!\!\!& Fscore \!\!\!\!& Fscore \hspace{-10pt}&\!\!\!\!& Fscore \!\!\!\!& Fscore \!\!\\
\wcline{1-6}
&\\[-9pt]
\textbf{[Strong labels]}&\\[0pt]
Conv. MTL \hspace{-2pt}\!\!&\!\! 87.64\% \!\!\!&\!\!\! 87.50\% \!\!\!&\!\!\!&\!\!\! \textbf{42.78\%} \!\!\!&\!\! 11.24\% \!\!\\
\textbf{[Weak labels]}&\\[0pt]
CNN {\small (ASC)} \hspace{-4pt}\!\!&\!\! 85.00\% \!\!\!&\!\!\! 84.29\% \!\!\!&\!\!\!&\!\!\! - \!\!\!&\!\! - \!\!\\[0pt]
CNN-BiGRU {\small (SED)} \hspace{-4pt}\!\!&\!\!\! - \!\!\!&\!\!\! - \!\!\!&\!\!\!&\!\!\! 30.71\% \!\!\!&\!\! 12.74\% \!\!\\[0pt]
MTL w/o MIL\hspace{-2pt}\!\!&\!\! 90.69\% \!\!\!&\!\!\! 90.80\% \!\!\!&\!\!\!&\!\!\! 32.41\% \!\!\!&\!\! 14.80\% \!\!\\
MTL w/ MIL {\small (MP)} \hspace{-2pt}\!\!&\!\! 89.94\% \!\!\!&\!\!\! 90.05\% \!\!\!&\!\!\!&\!\!\! 33.06\% \!\!\!&\!\! 15.04\% \!\!\\
MTL w/ MIL {\small (AP)} \hspace{-2pt}\!\!&\!\! 87.71\% \!\!\!&\!\!\! 87.70\% \!\!\!&\!\!\!&\!\!\! 32.47\% \!\!\!&\!\! 14.67\% \!\!\\
MTL w/ MIL {\small (ES)} \hspace{-2pt}\!\!&\!\! 90.22\% \!\!\!&\!\!\! 90.39\% \!\!\!&\!\!\!&\!\!\! 33.34\% \!\!\!&\!\! 15.04\% \!\!\\
MTL w/ MIL {\small (AT)} \hspace{-2pt}\!\!&\!\! \textbf{91.15\%} \!\!\!&\!\!\! \textbf{91.36\%} \!\!\!&\!\!\!&\!\!\! 33.09\% \!\!\!&\!\! \textbf{15.19\%} \!\!\\
\wcline{1-6}
\end{tabular}
\vspace{-5pt}
\end{table}
\begin{table}[t!]
\vspace{-4pt}
\small
\caption{Average Fscores for selected sound events}
\label{tab:performance02}
\hspace*{-5.5pt}
\centering
\begin{tabular}{lcccccc}
\wcline{1-6}\\[-9pt]
\!\!&\!\!{\bf  Bird}\!\!&\!\! \!\!&\!\! \!\!&\!\!{\bf Keyboard}\!\!&\!\!\!\!{\bf Large}\!\!\\[-1.5pt]
\multicolumn{1}{c}{\multirow{-1.9}{*}{\!\! \bf Method}}&\!\!{\bf  singing}\!\!&\!\!\multirow{-1.9}{*}{\!\bf  Car}\!\!&\!\!\!\multirow{-1.9}{*}{\!\bf Cutlery}\!\!\!&\!\!\!{\bf typing}\!\!\!&\!\!\!\!{\bf vehicle}\!\!\\[0.5pt]
\wcline{1-6}\\[-9pt]
\!\!Conv. MTL\!\!\!&\!\!46.29\%\!\!&\!\!45.51\%\!\!&\!\!0.00\%\!\!&\!\!\!5.08\%\!\!\!&\!\!\!\!12.29\%\!\!\\
\!\!CNN-BiGRU\!\!\!&\!\!39.64\%\!\!&\!\!44.86\%\!\!&\!\!\textbf{0.68\%}\!\!&\!\!\!6.96\%\!\!\!&\!\!\!\!11.37\%\!\!\\
\!\!MTL w/o MIL\!\!\!&\!\!51.67\%\!\!&\!\!46.28\%\!\!&\!\!0.20\%\!\!&\!\!\!11.66\%\!\!\!&\!\!\!\!14.44\%\!\!\\
\!\!MTL w/ MIL {\small (MP)}\!\!\!&\!\!\textbf{52.57\%}\!\!&\!\!\textbf{48.32\%}\!\!&\!\!0.21\%\!\!&\!\!\!\textbf{12.13\%}\!\!\!&\!\!\!\!14.59\%\!\!\\
\!\!MTL w/ MIL {\small (AP)}\!\!\!&\!\!49.22\%\!\!&\!\!47.36\%\!\!&\!\!0.10\%\!\!&\!\!\!10.66\%\!\!\!&\!\!\!\!14.79\%\!\!\\
\!\!MTL w/ MIL {\small (ES)}\!\!\!&\!\!51.48\%\!\!&\!\!47.55\%\!\!&\!\!0.03\%\!\!&\!\!\!11.36\%\!\!\!&\!\!\!\!\textbf{15.05\%}\!\!\\
\!\!MTL w/ MIL {\small (AT)}\!\!\!&\!\!52.53\%\!\!&\!\!47.14\%\!\!&\!\!0.43\%\!\!&\!\!\!11.50\%\!\!\!&\!\!\!\!14.61\%\!\!\\
\wcline{1-6}
\end{tabular}
\vspace{6pt}
\end{table}
%
%
%
\vspace{-7pt}
\section{Conclusions}
\label{sec:conclusion}
\vspace{-5pt}
In this paper, we proposed a method for the joint analysis of acoustic scenes and sound events using weak labels of sound events.
In the proposed method, we applied the MTL-based framework with the MIL based on max pooling, average pooling, exponential softmax pooling, and attention pooling.
The experimental results obtained using parts of the TUT Acoustic Scenes 2016/2017 and TUT Sound Events 2016/2017 datasets indicate that even when we use the weak labels of sound events, information on acoustic scenes and sound events improves SED and ASC, respectively.
Moreover, the experimental results also indicate that attention pooling can balance a reasonable performance between instantaneous and long-lasting sounds, and it achieves the best performance with the four pooling functions.
%
%
\vspace{-6pt}
\section{Acknowledgment}
\vspace{-4pt}
This work was supported by JSPS KAKENHI Grant Number JP20H00613.
%
%
\small
\bibliographystyle{IEEEbib}
\bibliography{IEEEabrv,KeisukeImoto12,IWAENC2022refs}

\begin{thebibliography}{10}

\bibitem{Fonseca_DCASE2018_01}
E.~Fonseca, M.~Plakal, F.~Font, D.~P.~W. Ellis, X.~Favory, J.~Jordi, and
  X.~Serra,
\newblock ``General-purpose tagging of freesound audio with {AudioSet} labels:
  Task description, dataset, and baseline,''
\newblock {\em Proc. Workshop on Detection and Classification of Acoustic
  Scenes and Events {\rm (}DCASE{\rm )}}, pp. 69--73, 2018.

\bibitem{Chan_EUSIPCO2010_01}
C.~Chan and E.~W.~M. Yu,
\newblock ``An abnormal sound detection and classification system for
  surveillance applications,''
\newblock {\em Proc. European Signal Processing Conference {\rm (}EUSIPCO{\rm
  )}}, pp. 1851--1855, 2010.

\bibitem{Koizumi_DCASE2020_01}
Y.~Koizumi, Y.~Kawaguchi, K.~Imoto, T.~Nakamura, Y.~Nikaido, R.~Tanabe,
  H.~Purohit, K.~Suefusa, T.~Endo, M.~Yasuda, and N.~Harada,
\newblock ``Description and discussion on {DCASE2020} challenge task2:
  Unsupervised anomalous sound detection for machine condition monitoring,''
\newblock {\em Proc. Detection and Classification of Acoustic Scenes and Events
  {\rm (}DCASE{\rm )}}, pp. 81--85, 2020.

\bibitem{Morfi_JASA2021_01}
V.~Morfi, R.~F. Lachlan, and D.~Stowell,
\newblock ``Deep perceptual embeddings for unlabelled animal sound,''
\newblock {\em {IEEE/ACM} Trans. Audio Speech Lang. Process.}, vol. 150, no. 1,
  pp. 2--11, 2020.

\bibitem{Valenti_IJCNN2017_01}
M.~Valenti, S.~Squartini, A.~Diment, G.~Parascandolo, and T.~Virtanen,
\newblock ``A convolutional neural network approach for acoustic scene
  classification,''
\newblock {\em Proc. International Joint Conference on Neural Networks {\rm
  (}IJCNN{\rm )}}, pp. 1547--1554, 2017.

\bibitem{Raveh_DCASE2018_01}
A.~Raveh and A.~Amar,
\newblock ``Multi-channel audio classification with neural network using
  scattering transform,''
\newblock {\em Tech. Rep. DCASE Challenge 2018 Task5}, pp. 1--4, 2018.

\bibitem{Cakir_TASLP2017_01}
E.~\c{C}ak\i r, G.~Parascandolo, T.~Heittola, H.~Huttunen, and T.~Virtanen,
\newblock ``Convolutional recurrent neural networks for polyphonic sound event
  detection,''
\newblock {\em {IEEE/ACM} Trans. Audio Speech Lang. Process.}, vol. 25, no. 6,
  pp. 1291--1303, 2017.

\bibitem{Kong_TASLP2020_01}
Q.~Kong, Y.~Xu, W.~Wang, and M.~D. Plumbley,
\newblock ``Sound event detection of weakly labelled data with
  {CNN}-{T}ransformer and automatic threshold optimization,''
\newblock {\em {IEEE/ACM} Trans. Audio Speech Lang. Process.}, vol. 28, pp.
  2450--2460, 2020.

\bibitem{Mesaros_EUSIPCO2011_01}
A.~Mesaros, T.~Heittola, and A.~Klapuri,
\newblock ``Latent semantic analysis in sound event detection,''
\newblock {\em Proc. European Signal Processing Conference {\rm (}EUSIPCO{\rm
  )}}, pp. 1307--1311, 2011.

\bibitem{Heittola_JASM2013_01}
T.~Heittola, A.~Mesaros, A.~Eronen, and T.~Virtanen,
\newblock ``Context-dependent sound event detection,''
\newblock {\em EURASIP Journal on Audio, Speech, and Music Processing}, vol.
  2013, no. 1, 2013.

\bibitem{Imoto_IEICE2016_01}
K.~Imoto and S.~Shimauchi,
\newblock ``Acoustic scene analysis based on hierarchical generative model of
  acoustic event sequence,''
\newblock {\em IEICE Transactions on Information and Systems}, vol. E99-D, no.
  10, pp. 2539--2549, 2016.

\bibitem{Imoto_TASLP2019_01}
K.~Imoto and N.~{Ono},
\newblock ``Acoustic topic model for scene analysis with intermittently missing
  observations,''
\newblock {\em {IEEE/ACM} Trans. Audio Speech Lang. Process.}, vol. 27, no. 2,
  pp. 367--382, 2019.

\bibitem{Bear_INTERSPEECH2019_01}
H.~L. Bear, I.~Nolasco, and E.~Benetos,
\newblock ``Towards joint sound scene and polyphonic sound event recognition,''
\newblock {\em INTERSPEECH}, pp. 4594--4598, 2019.

\bibitem{Tonami_WASPAA2019_01}
N.~Tonami, K.~Imoto, M.~Niitsuma, R.~Yamanishi, and Y.~Yamashita,
\newblock ``Joint analysis of acoustic events and scenes based on multitask
  learning,''
\newblock {\em Proc. {IEEE} Workshop on Applications of Signal Processing to
  Audio and Acoustics {\rm (}WASPAA{\rm )}}, pp. 333--337, 2019.

\bibitem{Jung_ICASSP2021_01}
J.~Jung, H.~J. Shim, J.~H. Kim, and H.~J. Yu,
\newblock ``{DCASENET}: An integrated pretrained deep neural network for
  detecting and classifying acoustic scenes and events,''
\newblock {\em Proc. {IEEE} International Conference on Acoustics, Speech and
  Signal Processing {\rm (}ICASSP{\rm )}}, pp. 621--625, 2021.

\bibitem{Kumar_ACMMM2016_01}
A.~Kumar and B.~Raj,
\newblock ``Audio event detection using weakly labeled data,''
\newblock {\em Proc. {ACM} International Conference on Multimedia {\rm
  (}ACMMM{\rm )}}, pp. 1038--1047, 2016.

\bibitem{Turpault_DCASE2019_01}
N.~Turpault, R.~Serizel, A.~Parag~Shah, and J.~Salamon,
\newblock ``{Sound Event Detection in Domestic Environments with Weakly Labeled
  Data and Soundscape Synthesis},''
\newblock {\em Proc. Workshop on Detection and Classification of Acoustic
  Scenes and Events {\rm (}DCASE{\rm )}}, pp. 253--257, 2019.

\bibitem{Zheng_DCASE2021_01}
X.~Zheng, H.~Chen, and Y.~Song,
\newblock ``Zheng {USTC} team’s submission for {DCASE2021} task4 -
  semi-supervised sound event detection,''
\newblock {\em Tech. Rep. DCASE Challenge 2021 Task4}, pp. 1--3, 2021.

\bibitem{Hershey_ICASSP2017_01}
S.~Hershey, S.~Chaudhuri, D.~P.~W. Ellis, J.~F. Gemmeke, A.~Jansen, R.~C.
  Moore, M.~Plakal, D.~Platt, R.~A. Saurous, B.~Seybold, M.~Slaney, R.~J.
  Weiss, and K.~Wilson,
\newblock ``{CNN} architectures for large-scale audio classification,''
\newblock {\em Proc. {IEEE} International Conference on Acoustics, Speech and
  Signal Processing {\rm (}ICASSP{\rm )}}, pp. 131--135, 2017.

\bibitem{Imoto_ICASSP2020_01}
K.~Imoto, N.~Tonami, Y.~Koizumi, M.~Yasuda, R.~Yamanishi, and Y.~Yamashita,
\newblock ``Sound event detection by multitask learning of sound events and
  scenes with soft scene labels,''
\newblock {\em Proc. {IEEE} International Conference on Acoustics, Speech and
  Signal Processing {\rm (}ICASSP{\rm )}}, pp. 621--625, 2020.

\bibitem{Dietterich_AI1997_01}
T.~G. Dietterich, R.~H. Lathrop, and T.~Lozano-P\'{e}rez,
\newblock ``Solving the multiple instance problem with axis-parallel
  rectangles,''
\newblock {\em Artificial Intelligence}, vol. 89, no. 1, pp. 31--71, 1997.

\bibitem{Wang_ICASSP2019_01}
Y.~Wang, J.~Li, and F.~Metze,
\newblock ``A comparison of five multiple instance learning pooling functions
  for sound event detection with weak labeling,''
\newblock {\em Proc. {IEEE} International Conference on Acoustics, Speech and
  Signal Processing {\rm (}ICASSP{\rm )}}, pp. 31--35, 2019.

\bibitem{Liu_ICLR2020_01}
L.~Liu, H.~Jiang, P.~He, W.~Chen, X.~Liu, J.~Gao, and J.~Han,
\newblock ``On the variance of the adaptive learning rate and beyond,''
\newblock {\em Proc. International Conference on Learning Representations {\rm
  (}ICLR{\rm )}}, pp. 1--13, 2020.

\bibitem{Mesaros_EUSIPCO2016_01}
A.~Mesaros, T.~Heittola, and T.~Virtanen,
\newblock ``{TUT} database for acoustic scene classification and sound event
  detection,''
\newblock {\em Proc. European Signal Processing Conference {\rm (}EUSIPCO{\rm
  )}}, pp. 1128--1132, 2016.

\bibitem{Mesaros_DCASE2017_01}
A.~Mesaros, T.~Heittola, A.~Diment, B.~Elizalde, A.~Shah, B.~Raj, and
  T.~Virtanen,
\newblock ``{DCASE} 2017 challenge setup: Tasks, datasets and baseline
  system,''
\newblock {\em Proc. Workshop on Detection and Classification of Acoustic
  Scenes and Events {\rm (}DCASE{\rm )}}, pp. 85--92, 2017.

\bibitem{Imoto_dataset2019_01}
\url{https://www.ksuke.net/dataset}.

\bibitem{Imoto_ICASSP2021_01}
K.~Imoto, S.~Mishima, Y.~Arai, and R.~Kondo,
\newblock ``Impact of sound duration and inactive frames on sound event
  detection performance,''
\newblock {\em Proc. {IEEE} International Conference on Acoustics, Speech and
  Signal Processing {\rm (}ICASSP{\rm )}}, pp. 875--879, 2021.

\end{thebibliography}

\end{document}